\documentclass[aps, pre, showpacs, showkeys, twocolumn]{revtex4-1}

\usepackage{amsmath,amssymb,amsthm,amsfonts,latexsym}
\usepackage{bm,amsfonts, mathtools}
\usepackage{graphicx,color, wasysym}
\usepackage{textcomp}
\usepackage{amsmath,amssymb,latexsym,epsfig}
\usepackage{appendix}

\def\ulamek#1#2{\mbox{\normalfont$\frac{#1}{#2}$}}

\DeclareMathOperator{\I}{i}

\begin{document}

\title[The Havriliak-Negami relaxation and its relatives: the response, relaxation and probability density functions]{The Havriliak-Negami relaxation and its relatives: the response, relaxation and probability density functions}

\author{K.~G\'{o}rska} 
\email{katarzyna.gorska@ifj.edu.pl}

\author{A.~Horzela}
\email{andrzej.horzela@ifj.edu.pl}

\author{{\L}.~Bratek}
\email{lukasz.bratek@ifj.edu.pl}
\affiliation{H. Niewodnicza\'{n}ski Institute of Nuclear Physics, Polish Academy of Sciences, Division of Theoretical Physics, ul. Eliasza-Radzikowskiego 152, PL 31342 Krak\'{o}w, Poland}

%
%
\author{K.~A.~Penson}
\email{penson@lptl.jussieu.fr}
\affiliation{Sorbonne Universit\'{e}s, Universit\'{e} Pierre et Marie Curie (Paris 06), CNRS UMR 7600, Laboratoire de Physique Th\'{e}orique de la Mati\`{e}re Condens\'{e}e (LPTMC), Tour 13-5i\`{e}me \'{e}t., B.C. 121, 4 pl. Jussieu, F 75252 Paris Cedex 05, France}

\author{G. Dattoli}
\email{giuseppe.dattoli@enea.it}
\affiliation{ENEA - Centro Ricerche Frascati, via E. Fermi, 45, IT 00044 Frascati (Roma), Italy}

\begin{abstract}
We study functions related to the experimentally observed Havriliak-Negami dielectric relaxation pattern in the frequency domain $\sim [1+(\I\omega\tau_{0})^{\alpha}]^{-\beta}$ with $\tau_{0}$ being some characteristic time. For $\alpha = l/k< 1$  ($l$ and $k$ positive integers) and $\beta > 0$ we furnish exact and explicit expressions for response and relaxation functions 
in the time domain and suitable probability densities 
in their "dual" domain. All these functions are expressed as finite sums of generalized hypergeometric functions, convenient to handle analytically and numerically.
Introducing a reparameterization $\beta = (2-q)/(q-1)$ and $\tau_{0} = (q-1)^{1/\alpha}$ $(1 < q < 2)$ we show that for $0 < \alpha < 1$ the response functions $f_{\alpha, \beta}(t/\tau_{0})$  go to the one-sided L\'{e}vy stable distributions when $q$ tends to one. Moreover, applying the self-similarity property of the probability densities $g_{\alpha, \beta}(u)$, we introduce two-variable densities and show that they satisfy the integral form of the evolution equation. 
\end{abstract}



\maketitle

\section{Introduction}

In many glass-forming systems, {like amorphous polymers or supercooled liquids near the glass transition temperature, the relaxation spectrum exhibits strongly non-exponential behavior. Dielectric spectroscopy shows evidence of two relaxation processes called the $\alpha-$ and $\beta$-relaxations. The $\alpha$-relaxation is described by an asymmetric peak which flattens into the $\beta$-relaxation at high frequency domain \cite{RHilfer02b}.} It is assumed that the $\alpha$-relaxation corresponds to the motion of the clusters of atoms or to the atomic transport between clusters, while the $\beta$-relaxation corresponds to some motion of atoms within the clusters. In a phenomenological approach these two types of relaxation are usually described by the sum of empirical non-Debye laws, namely the sum of Havriliak-Negami (HN) \cite{SHavriliak67} relaxation functions \cite{PRosner04, GKatana95, SCerveny08, ADoss02, MGalazka15} or their sums with the so-called stretched exponentials \cite{RCasalini03, KFukao00, RCasalini11}, named also the Kohlrausch-Williams-Watts (KWW) functions. The latter were extensively studied in many theoretical (see e.g. \cite{VVUchaikin99, BDHughes95, ELukacs70, KAPenson10}) and experimental papers (see e.g.\cite{LCipelletti05, JSabelko99, JBredenbeck05, GDattoli14, VSaltas07}). The frequency dependent behavior  of the dielectric polarizability was also described by the excess wing model \cite{RHilfer02b, RHilfer02c} or by the Kilbas-Saigo function \cite{ECapelasDeOliveira14}.

The HN function, introduced in \cite{SHavriliak67} to describe the frequency dependence  of the absolute dielectric permittivity $\hat{\varepsilon}(\omega)$, is conventionally written down as
\begin{equation}\label{3/06-2}
\frac{\hat{\varepsilon}(\omega) - \varepsilon_{\infty}}{\varepsilon - \varepsilon_{\infty}} =  \frac{1}{[1 + (i\omega\tau_{0})^{\alpha}]^{\beta}}, \quad 0 < \alpha < 1, \quad \beta > 0,
\end{equation}
where $\alpha$ and $\beta$, called respectively the width and the symmetry parameters, are experimentally fitted numbers having nothing in common with the previously mentioned  $\alpha-$ and $\beta$-relaxations. In Eq. \eqref{3/06-2} $\omega$ is the frequency, $\tau_{0}$ means an effective time constant while the symbols $\varepsilon$ and $\varepsilon_{\infty}$  denote the relative permittivity and the dielectric constant. 
Eq. \eqref{3/06-2} generalizes the Cole-Cole (CC) \cite{CJFBottcher78} relaxation (Eq. \eqref{3/06-2} for $\beta=1$) and the Cole-Davidson (CD) \cite{CJFBottcher78} relaxation (Eq. \eqref{3/06-2} for $\alpha=1$). The use of the HN functions is the standard method to describe experimental data for a very large range of different phenomena. We mention \cite{VSaltas07} where the HN function was applied for monitoring the contamination in sandstone, and investigations of complex systems representing plant tissues of fresh fruits and vegetables for which the HN relaxation in the frequency range $10^{7}-1.8 \times10^{9}$ Hz has appeared to be an useful tool of analysis \cite{RRNigmatulin06}. 

Universality of the HN relaxation pattern motivated several authors who have investigated it from different points of view. For example, the mutual relation between CC relaxation and the stretched exponential was established in  \cite{KWeron96, RMetzler02}. A transparent subordination approach to anomalous diffusion processes underlying the HN relaxation was proposed in \cite{KWeron10, AStanislavsky15}. The HN relaxation was also obtained from the generalization of the standard continuous-time random walk, called the coupled memory continuous-time random walk \cite{EGudowskaNowak05}. In more mathematically oriented studies it was found that analytic expressions for the HN relaxation in the time domain are expressed for real values of $\alpha$ and $\beta$ in terms of the Fox $H$ functions \cite{RHilfer02, RHilfer02a}. We consider this result very important from the theoretical point of view but it is a little impractical for experimentalists as expressions involving the Fox $H$ functions for general values of parameters are purely symbolic (in fact the Fox $H$ functions are defined via the Mellin transform \cite{APPrudnikov-v3}) and the Fox $H$ functions themselves are not implemented in the standard computer algebra systems. 

The purpose of this paper is twofold. First, we are going to show how to express the functions relevant to the HN relaxation, like the response and relaxation functions and the probability densities, in terms of the special functions implemented in the computer algebra systems. That significantly simplifies calculations and makes them accessible for much larger community. The crucial step to realize this purpose is that we replace the real values of the parameter $\alpha$ by the rationals such that $0 < \alpha < 1$. Hence, everywhere in what follows we shall take $\alpha = l/k$ with $l$ and $k$  being relative prime integers. Our second aim is to propose how to calculate the probability density related to the given response function. The knowledge of such a quantity should allow one to get an additional information on the properties of relaxing centers located within a given sample.

The paper is organized as follows. Sec. II contains some facts about the relaxation processes, namely the relation between the response functions, the probability densities and the relaxation functions. In Sec. III, the response functions $f_{\alpha, \beta}(t/\tau_{0})$, $0 < \alpha < 1$ and $\beta >0$, are represented via the Prabhakar function which contains the (three-parameter) generalized Mittag-Leffler functions. For $\alpha = l/k < 1$, where $l$ and $k$ are relatively prime integers, we represent the (three-parameter) generalized Mittag-Leffler functions as finite sums of the generalized hypergeometric functions. Next, we find the asymptotic properties of the (three-parameter) generalized Mittag-Leffler functions and, consequently, the response functions. Expressing the response function in terms of the Meijer $G$ function  we calculate, in Sec. IV, the probability density $g_{l/k, \beta}(u)$ connected with the HN relaxation. We show that such obtained  $g_{l/k, \beta}(u)$ is positively defined and normalized for $\beta < k/l$. We also find all the moments and asymptotics of $g_{\alpha, \beta}(u)$. In Sec. V the relaxation functions related to the HN function, denoted there as $[n(t)/n_{0}]_{\alpha, \beta}$, are derived and the integral form of the evolution equation is found. The new relation between $f_{\alpha, \beta}(t/\tau_{0})$ and the one-sided L\'{e}vy stable densities is proposed in Sec. VI. The paper is concluded in Sec. VII.

\section{Some remarks on the relaxation theory}

In the relaxation theory the complex frequency-dependent absolute permittivity (dielectric "constant") of the dielectric medium is given by \cite{CJFBottcher78}
\begin{equation}\label{3/06-3}
\frac{\hat{\varepsilon}(\omega) - \varepsilon_{\infty}}{\varepsilon - \varepsilon_{\infty}} = \mathcal{L}[f(t/\tau_0); i\omega], 
\end{equation}
where $\mathcal{L}[f(t/\tau_0); p] = \int_{0}^{\infty} e^{-p t} f(t/\tau_0) dt$ 
denotes the Laplace transform \cite{INSneddon74}  of $f(t/\tau_0)= -\frac{d}{d t} \frac{n(t)}{n_{0}}$. The latter connects the responce function $f(t/\tau_{0})$ and $\frac{n(t)}{n_{0}}$, the ratio of instance number of polarized centers taken at a moment of time $t$ to their initial number. Inserting this relation into 
Eq. \eqref{3/06-3}, comparing it with Eq. \eqref{3/06-2}, and using Eq. (3-4-1) of \cite{INSneddon74} we get 
\begin{equation}\label{5/09-1}
[1+(i\omega\tau_{0})^{\alpha}]^{-\beta} = 1 -i\omega\, \mathcal{L}\left[\frac{n(t)}{n_{0}}; i\omega\right].
\end{equation}
Extracting the Laplace transform from Eq. \eqref{5/09-1} and, thereafter, inverting it R. Hilfer calculated in \cite{RHilfer02, RHilfer02a} all standard non-Debye relaxation functions $n(t)/n_{0}$, in general expressing them through the Fox $H$ functions. 

Adopting the other point of view one may consider $n(t)/n_{0}$ as a number of initially polarized centers which did not relax during the time interval $(0, t)$. If we assume that $N$ polarized centers relax according to the Debye laws with a different characteristic relaxation times $\tau_{k} = \tau_{0}/u_{k}$, $k=1,\dots,N$ then the relaxation of the sample, treated as a whole,  should be a weighted sum of the Debye relaxations, namely 
\begin{equation*}
\frac{n(t)}{n_{0}} = \sum_{k} e^{- t \ulamek{u_{k}}{\tau_{0}}} g(u_{k}) \Delta u_{k}, 
\end{equation*}
where the probability density $g(u_{k})$ is a positive function which satisfies normalization condition $\sum_{k} g(u_{k}) \Delta u_{k} = 1$. Going with $N$ to infinity, equivalent to take infinitesimally small $\Delta u_{k}$, we get
\begin{align}\label{4/06-1}
\frac{n(t)}{n_{0}} & = \int_{0}^{\infty} e^{-\ulamek{t}{\tau_{0}} u} g(u) du = \int_{0}^{\infty} e^{-v} g(\ulamek{\tau_{0}}{t}v) \ulamek{\tau_{0}}{t} dv,
\end{align}
where $v = t u/\tau_{0}$. Taking the time derivative of the first equality in Eq. \eqref{4/06-1} provides us with the expression for the response function
\begin{equation}\label{4/06-5}
f(t) = \int_{0}^{\infty} e^{-\ulamek{t}{\tau_{0}} u} \ulamek{u}{\tau_{0}} g(u) du,
\end{equation}
which means that the response function is somehow built from the Debye relaxation. 
Assuming the equality between expressions for $n(t)/n_{0}$ arising from the Eqs. \eqref{5/09-1} and \eqref{4/06-1} one concludes that the function $g(u)$ may be written down as an inverse Laplace transform of the Fox $H$ function. 

\section{The response function}

Consider the HN relaxation and denote the related to it response function in the time domain as $f(t/\tau_{0})~\equiv~f_{\alpha, \beta}(t/\tau_{0})$ where $\alpha$ and $\beta$ are the width and symmetry parameters introduced in Eq. \eqref{3/06-2}. To calculate the function $f_{\alpha, \beta}(t/\tau_{0})$ we compare Eqs. \eqref{3/06-2} and \eqref{3/06-3}, and, then, invert the Laplace transform. Using Eq. (2.5) of \cite{TRPrabhakar71} for $\beta_{\rm P} = \alpha\beta$, $\rho_{\rm P} = \beta$ and $\lambda_{\rm P} = -1$ (the subscript $\rm P$ is added to emphasize the reference from which the formula is taken) we obtain the response function expressed via the so-called Prabhakar function  \cite{TRPrabhakar71, FMainardi15, TKPogany16, Lav}:
\begin{equation}\label{5/09-2}
f_{\alpha, \beta}(\ulamek{t}{\tau_{0}}) = \frac{1}{\tau_{0}}\, \big(\!\ulamek{t}{\tau_{0}}\!\big)^{\alpha\beta - 1} E^{\beta}_{\alpha, \alpha\beta}(-\big(\!\ulamek{t}{\tau_{0}}\!\big)^{\alpha}),
\end{equation}
where $E^{\delta}_{\alpha, \gamma}(z)$ is the (three-parameter) generalized Mittag-Leffler function given by the series \cite{FMainardi15, ECapelasDeOliveira11, TKPogany16, AAKilbas}
\begin{equation}\label{5/09-3}
E^{\delta}_{\alpha, \gamma}(z) = \sum_{n=0}^{\infty}\frac{(\delta)_{n}  z^{n}}{n! \Gamma(\alpha n + \gamma)},
\end{equation} 
where $(\delta)_{n} = \Gamma(\delta+n)/\Gamma(\delta)$ denotes the Pochhammer symbol. Here we point out that there exists in the literature another definition of the three-parameter Mittag-Leffler function \cite{ECapelasDeOliveira14}, but it will be neither used nor discussed by us here. The (three-parameter) generalized Mittag-Leffler function used in what follows is much less known than its one- and two-parameter analogues provided by special choices of parameters, e.g. Eq. \eqref{5/09-3} for $\delta = 1$  gives $E^{1}_{\alpha, \gamma}(z) = E_{\alpha, \gamma}(z)$ which is the (two-parameter) Mittag-Leffler function \cite{AAKilbas}. Another particular case of Eq. \eqref{5/09-3}, namely taking  $\delta = \gamma = 1$,  leads to the classical (one-parameter) Mittag-Leffler function $E_{\alpha}(z)$ \cite{AAKilbas} while the choice  $\delta = 1$ and $\gamma = 1 + \alpha$ gives $z E^{1}_{\alpha, 1+\alpha}(z) = E_{\alpha}(z) - 1$. Numerical calculations of $E_{\alpha, \gamma}(z)$ were performed in \cite{HSeybold08, RHilfer06}, where the authors constructed and successfully used the algorithm based on integral representations and exponential asymptotics. Additionally they tested stability of their algorithm and validity range for the parameters $\alpha$ and $\gamma$. We would like also to remind that the authors of \cite{HSeybold08, RHilfer06} simulated the asymptotic behavior of $E_{\alpha, \gamma}(z)$ at zero and infinity, and estimated the error of their method.

As mentioned in the above the (three-parameter) generalized Mittag-Leffler functions of Eq. \eqref{5/09-3} are not widely recognized objects and our goal in what follows is to express them as finite sums of much better known generalized hypergeometric functions ${_{p}F_{q}}$ \cite{APPrudnikov-v3}. For this reason we assume $\alpha$ to be rational $0 < \alpha = l/k < 1$, and change in Eq. \eqref{5/09-3} the {sum} over $n = 0, 1, \ldots$ into double {sum} over $n'=kn$ and $j = 0, 1, \ldots, k-1$.
{The sum} over $n'$ leads to the generalized hypergeometric functions 
while {the sum} over $j$ indicates how many hypergeometric functions do appear. Thus, after some algebraic manipulations we get
\begin{align}\label{4/06-9}
\begin{split}
& E^{\delta}_{l/k, \gamma}(z) = \sum_{j=0}^{k-1} \frac{z^{j}}{j!} \frac{(\delta)_{j}}{\Gamma(\gamma + \frac{l}{k}j)} \\
& \qquad \times {_{1+k}F_{l+k}}\left({1, \Delta(k, \delta+j) \atop \Delta(k, 1+j), \Delta(l, \gamma + \frac{l}{k}j)}; \frac{z^{k}}{l^{l}}\right)
\end{split}
\end{align}
with $\Delta(n, a) = \ulamek{a}{n}, \ulamek{a+1}{n}, \ldots, \ulamek{a+n-1}{n}$.  The list of "upper" parameters includes $1$ followed by $\Delta(k, \delta~+~j)$, whereas the list of "lower" parameters is the union of $\Delta(k, 1+j)$ and $\Delta(l, \gamma + \frac{l}{k}j)$.  {To check if Eq. \eqref{4/06-9} reconstructs the exponential behavior of $E_{\alpha, \gamma}(z)$ obtained in \cite{HSeybold08, RHilfer06} we estimate $E^{\delta}_{l/k, \gamma}(z)$ for large values of $z$. To begin with we consider Eq. \eqref{4/06-9} and the last unnumbered formula on p. 155 of \cite{EWBarnes07}. According to it and the Gauss-Legendre multiplication formula for $\Gamma$ functions the generalized hypergeometric function ${_{1+k}F_{l+k}}$ is asymptotically equal to $[lj! \Gamma(\gamma + lj/k)/\Gamma(\delta + j)] (k/l)^{\delta-1} \exp(z^{k/l}) z^{k(\delta-\gamma)/l - j}$ as $z\to \infty$. Substituting this relation to Eq. \eqref{4/06-9} we get
\begin{equation*}
E^{\delta}_{l/k, \gamma}(z) \sim \frac{(k/l)^{\delta}}{k\Gamma(\delta)} z^{k(\delta - \gamma)/l} e^{z^{k/l}}, \quad z\gg 1,
\end{equation*}
which for $\alpha = l/k$ and $\delta = 1$ reproduces the leading term of \cite[Eq. (2.4)]{HSeybold08}, i.e. the exponential behavior of $E_{l/k, \gamma}$.} 

Eqs. \eqref{5/09-2} and \eqref{4/06-9} with $\delta = \beta$ and $\gamma = \alpha\beta$ put in mean that $f_{l/k, \beta}(t/\tau_{0})$ may be represented as a finite sum of ${_{1+k}F_{l+k}}$'s functions,  for $k = 2$ simplifying to 
\begin{equation}\label{26/09-1}
f_{1/2, \beta}(\ulamek{t}{\tau_{0}}) = \frac{\sqrt{2}\tau_{0}\beta}{\sqrt{\pi}} \left(\frac{2t}{\tau_{0}}\right)^{\frac{\beta}{2}-1}\!\! e^{\frac{t}{2\tau_{0}}} D_{-1-\beta}\big(\!\sqrt{\ulamek{2t}{\tau_{0}}}\big),
\end{equation}
where $D_{\nu}(z)$ is the parabolic cylinder function \cite{Gradshteyn07}. In the derivation of Eq. \eqref{26/09-1} we have employed Eqs. (7.11.1.9) and (7.11.1.10) on p. 579 of \cite{APPrudnikov-v3}.

We would like also to put the readers' attention to the fact that for $\delta = \beta$ and $\gamma = \alpha\beta$ $E^{\delta}_{\alpha, \gamma}(z)$ can be represented via the one-sided L\'{e}vy stable distribution $\varPhi_{\alpha}(z)$ \cite{KAPenson10}:
\begin{equation}\label{4/06-7}
E^{\beta}_{\alpha, \alpha\beta}(z) = \int_{0}^{\infty} e^{z s} \frac{s^{\beta-(1+\frac{1}{\alpha})}}{\Gamma(\beta)} \varPhi_{\alpha}(s^{-\frac{1}{\alpha}}) ds.
\end{equation}
One can demonstrate Eq. \eqref{4/06-7} if takes the series expansion of the exponential function and subsequently uses the explicitly known values of moments of $\varPhi_{\alpha}(u)$ (see \cite{VVUchaikin99, KAPenson10}, consult also \cite{GP11, GP12, PG16, TPog} for closely related algebraic and transformation properties of L\'{e}vy stable functions). Note that Eq. \eqref{4/06-7} generalizes the known relation between the classical Mittag-Leffler function and the one-sided L\'{e}vy stable distribution (\cite[Eq. (7)]{KWeron96} or  \cite[Eq. (11)]{KGorska12}). Inserting into Eq. \eqref{4/06-7} the large $u$ asymptotics of $\varPhi_{\alpha}(u)\simeq u^{-1-\alpha}$ (given by \cite[Eq. (5)]{JMikusinski59}) we get $E^{\beta}_{\alpha, \alpha\beta}(-(t/\tau_{0}))^{\alpha})\simeq (t/\tau_{0})^{-\alpha-\alpha\beta}$. Substituting this result into Eq. \eqref{5/09-2} we reconstruct the asymptotics given in the last unnumbered formula on p. 76 of \cite{FMainardi15}, namely $f_{\alpha, \beta}(t/\tau_{0}) \simeq (t/\tau_{0})^{-1-\alpha}$ for $t\to\infty$. We would like to emphasize  that the behavior of $f_{\alpha, \beta}(t/\tau_{0})$ for $t\gg 1$ looks similarly to the asymptotics of $\varPhi_{\alpha}(u)$ for $u\gg 1$. In our opinion it suggests the existence of deeper relation between the Prabhakar function and the one-sided L\'{e}vy stable distributions. We shall come back to this problem in  Sec. VI and investigate it in a more detailed way. 

We complete this section providing results which will simplify calculations presented in Sec. IV. For this purpose we insert Eq. \eqref{4/06-7} with $\varPhi_{\alpha}(z)$ given by \cite[Eq.~(2)]{KAPenson10} into Eq. \eqref{5/09-2} and employ formula (2.24.3.1) on p. 295 of \cite{APPrudnikov-v3}. This implies that the Prabhakar function can be given in terms of the Meijer $G$ function $G^{m, n}_{p, q}$ \cite{APPrudnikov-v3}:
\begin{align}\label{4/06-4}
\begin{split}
f_{l/k, \beta}(\!\ulamek{t}{\tau_{0}}\!) &= (2\pi)^{\ulamek{1+l}{2} - k} \frac{\sqrt{l} k^{\beta}}{\Gamma(\beta) t}\\
&\times G^{k, k}_{l+k, k}\left(\frac{l^{l}\tau_{0}^{l}}{t^{l}}\Big\vert{\Delta(k, 1-\beta), \Delta(l, 0) \atop \Delta(k, 0)}\right)
\end{split}
\end{align}
with the upper and lower lists of parameters equal to the union of $\Delta(k,~1~-~\beta)$ and $\Delta(l, 0)$ and $\Delta(k, 0)$, respectively. Eq. \eqref{4/06-4} with Eq. (2.24.2.1) on p. 293 of \cite{APPrudnikov-v3} used enable us to calculate the $\mu$-th Stieltjes moments of $f_{l/k, \beta}(\!\ulamek{t}{\tau_{0}}\!)$:
\begin{equation}\label{25/09-1}
{M}_{l/k, \beta}(\mu) = \int_{0}^{\infty} t^{\mu} f_{l/k, \beta}(\ulamek{t}{\tau_{0}}) dt=\tau_{0}^{\mu}\frac{\Gamma(\beta + \ulamek{k \mu}{l})}{\Gamma(\beta)} \tilde{{M}}_{l/k}(\mu), 
\end{equation}
where $\tilde{{M}}_{l/k}(\mu) = k\Gamma(-k\mu/l)/[l\Gamma(-\mu)]$ denote the $\mu$-th Stieltjes moments of the one-sided L\'{e}vy stable distribution \cite{KAPenson10}. The moments ${M}_{l/k, \beta}(\mu)$ are finite for $-l\beta/k < \mu < l/k$ and infinite otherwise, i.e. they exist in the narrower range than the $\tilde{{M}}_{l/k}(\mu)$ which are finite for the broader range of $\mu$, namely for $-\infty < \mu < l/k$.

\begin{figure}[!h]
\includegraphics[scale=0.38]{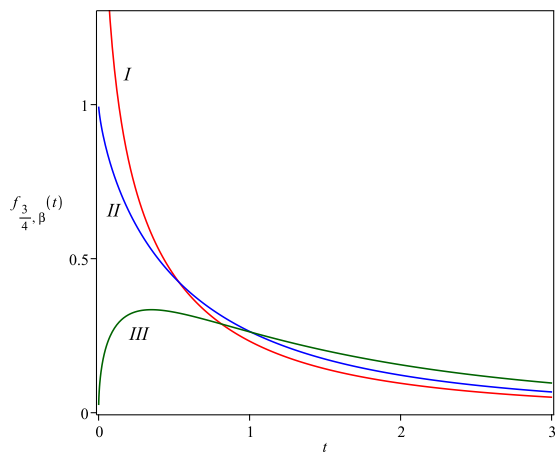}
\caption{\label{fig0}(Color online) Plot of $f_{l/k, \beta}(t)$ given by Eq. \eqref{4/06-4} for $\tau_{0}=1$, $\alpha = 3/4$ and $\beta = 1$ (I; red), $\beta = 4/3$ (II; blue), and $\beta = 2$ (III; green).} 
\end{figure}
Eq. \eqref{4/06-4} can be readily used to study the Prabhakar functions graphically. As an example we consider $f_{3/4, \beta}(t)$ 
for $\beta = 1, 3/4, 2$  plotted in the Fig. \ref{fig0}. In the limit $t\to 0$ the function $f_{3/4, \beta}(t)$  goes to infinity for $\beta < 4/3$, approaches 1 for $\beta = 4/3$ and vanishes for $\beta > 4/3$. That exemplifies and clarifies the known asymptotic behavior of $f_{\alpha, \beta}(t) \propto t^{\alpha\beta-1}$ shown to go to $1$ for $t\to 0$ if $\beta = 1/\alpha$ \cite{BSzabat07}. It also shows that $f_{3/4, \beta}(t)$ vanishes when $t$ tends to infinity which confirms the asymptotics described previously below Eq. \eqref{4/06-7}.  

\section{The probability density}

The inverse Laplace transform allows one to pass from the time domain $t>0$ to the "dual" domain $u>0$ and to address the question of emergence of probability distribution functions (pdf) in $u$.

\begin{figure}[!h]
\includegraphics[scale=0.38]{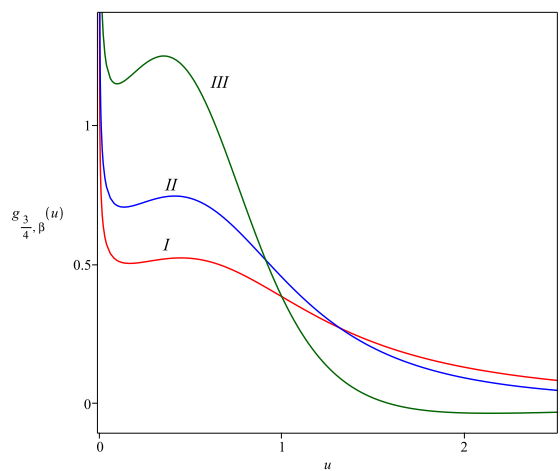}
\caption{\label{fig0a}(Color online) Plot of $g_{l/k, \beta}(u)$ given by Eq. \eqref{5/06-1} for $\alpha = 3/4$ and $\beta = 1$ (I; red), $\beta = 4/3$ (II; blue), and $\beta = 2$ (III; green). Observe that $g_{3/4, 2}(u)$ (line III (green)) becomes negative for $u > 1.33$.}
\end{figure}

To emphasize the fact that we are investigating the HN relaxation, from now on the function $g(u)$ of the Sec. II will be denoted as $g_{\alpha, \beta}(u)$. For the case under consideration, i.e.  $\alpha~=~l/k$, the inverse Laplace transform of Eq. \eqref{4/06-5} can be written down as:
\begin{equation}\label{13/09-1}
g_{l/k, \beta}(u) = \frac{\tau_{0}}{u} \mathcal{L}^{-1}[f_{l/k, \beta}(\!\ulamek{t}{\tau_{0}}\!), u]. 
\end{equation}
Substituting the Prabhakar function given by Eq. \eqref{4/06-4} and employing Eq. (3.38.1.2) on p. 393  of \cite{APPrudnikov-v5} we conclude that $g_{l/k, \beta}(u)$ is representable by the Meijer $G$ function \cite{APPrudnikov-v3}:
\begin{equation}\label{5/06-1}
g_{l/k, \beta}(u) = \frac{(2 \pi)^{l-k} k^{\beta}}{\Gamma(\beta) u} G^{k, k}_{l+k, l+k}\left(u^{l}\Big\vert{\Delta(k, 1-\beta), \Delta(l, 0) \atop \Delta(k, 0), \Delta(l, 0)}\right),
\end{equation}
with the symbols $\Delta(n, a)$ defined as below Eq. \eqref{4/06-9}. Numerical tests (see Fig. \ref{fig0a}) show that $g_{l/k, \beta}(u)$ are positive for $0 < \beta \leq k/l$, whereas they have negative parts for $\beta > k/l$. This conjecture will be confirmed analytically at the end of this Section, namely below the Eq. \eqref{14/06-2}.
Employing Eq. (2.24.2.1) on p. 293 of \cite{APPrudnikov-v3} we show that fractional moments $N_{l/k, \beta}(\nu) = \int_{0}^{\infty} u^{\nu} g_{l/k, \beta}(u) du$ are equal to
\begin{equation*}
N_{l/k, \beta}(\nu) = \frac{\Gamma(1 + \ulamek{k}{l}\nu)\Gamma(\beta - \ulamek{k}{l}\nu)}{\Gamma(\beta)\Gamma(1+\nu)\Gamma(1-\nu)}
\end{equation*}
which {are} finite for real $\nu$, $-\infty < \nu < l\beta/k$ and $\nu \neq -1, -2, \ldots$, while infinite otherwise. It means that $g_{l/k, \beta}(u)$ are normalized but their higher moments, e.g. the mean value or variance, do not exist. Similar (but not identical) behavior is observed for the one-sided L\'{e}vy stable distributions $\varPhi_{\alpha}(u)$ which fractional moments exist but higher moments are infinite \cite{KAPenson10}. Like in the case of the L\'{e}vy distributions the normalizability and positivity of $g_{l/k, \beta}(u)$ for $0 < \beta \leq k/l$ permits us to identify them as pdfs, but in contradistinction to the L\'{e}vy distributions 
the functions $g_{l/k, \beta}(u)$ are not necessarily unimodal.

Using Eq. (8.2.2.4) on p. 617 of \cite{APPrudnikov-v3} and the Gauss-Legendre multiplication formula for $\Gamma$ functions in Eq. \eqref{5/06-1}, we can rewrite $g_{l/k, \beta}(u)$ as a finite sum of $k$ generalized hypergeometric functions:
\begin{align}\label{5/06-2}
\begin{split}
g_{l/k, \beta}(u) & = \frac{1}{\pi} \sum_{j=0}^{k-1} \frac{(-1)^{j}}{j!} \frac{(\beta)_{j}}{u^{1+\frac{l}{k}(\beta+j)}} \sin[\ulamek{l}{k}(\beta+j)\pi]  \\
& \times {_{1+k}F_{k}}\left({1, \Delta(k, \beta+j) \atop \Delta(k, 1+j)}; \frac{(-1)^{l-k}}{u^{l}}\right).
\end{split}
\end{align} 
Eq. \eqref{5/06-2} gives a closed form of $g_{{l}/{k}, \beta}(u)$ which is convenient and efficiently applicable in calculations, both done by hand as well as using the standard computer algebra systems. For example, for the CC relaxation ($\beta = 1$) it is seen that, because of appropriate cancellations,  ${_{1+k}F_{k}}$'s give ${_{1}F_{0}}\left({1 \atop 0}, (-1)^{l-k}u^{-l/k}\right)$ which, after using Eq. (7.3.1.1) on p. 453 of \cite{APPrudnikov-v3}, yields $u^{l}/[u^{l} - (-1)^{l-k}]$. Employing it and Eq. (1.353.1) on p. 38 of \cite{Gradshteyn07} to the sum over $j$ we get 
\begin{equation}\label{16/06-1}
g_{\alpha, 1}(u) = \frac{u^{\alpha -1} \sin(\alpha\pi)}{\pi (u^{2\alpha} + 2 u^{\alpha} \cos(\alpha\pi) + 1)},
\end{equation}
with $0 <\alpha = l/k < 1$. We point out that Eq. \eqref{16/06-1} was obtained in \cite{RGorenflo08, ECapelasDeOliveira14, BDybiec10} using different methods, see Eq. (3.24) on p. 245 in \cite{RGorenflo08}, Eqs. (22) and (39) in \cite{ECapelasDeOliveira14} or Eq. (26) in \cite{BDybiec10}. Distributions  $g_{\alpha, 1}(u)$ share the following properties: (i) $g_{\alpha, 1}(u)$ are positive for $0 < \alpha < 1$ and $u \geq 0$; (ii) $g_{\alpha, 1}(u)$ go to infinity at $u=0$, and vanish for $u\to\infty$; (iii) the shape of $g_{\alpha, 1}(u)$ depends on the current value of $\alpha$: there exists a value $\alpha_0$ such that for $\alpha < \alpha_{0}$ $g_{\alpha, 1}(u)$ is monotonically decreasing function of $u$, while it has a kind of plateau if  $\alpha$ is close to $\alpha_{0}$ and possesses two distinct extrema for $\alpha > \alpha_{0}$. For the CC relaxation we have $\alpha_{0} \approx 0.737$ and the derivative $g'_{\alpha_0, 1}(u)$ vanish at a point $u_{0}\approx 0.306$ while for $\alpha > \alpha_{0}$ $g_{\alpha, 1}(u)$ has a maximum at $u = u_{\rm max}$ and a minimum at $u = u_{\rm min}$ where $u_{\rm max/ min} = [-\cos(\alpha\pi)\pm (\alpha^{2} - \sin^{2}(\alpha \pi))^{1/2}]/(1+\alpha)$ with the upper/lower signs applied for $u_{\rm max/ min}$, respectively.

\begin{figure}[!h]
\includegraphics[scale=0.40]{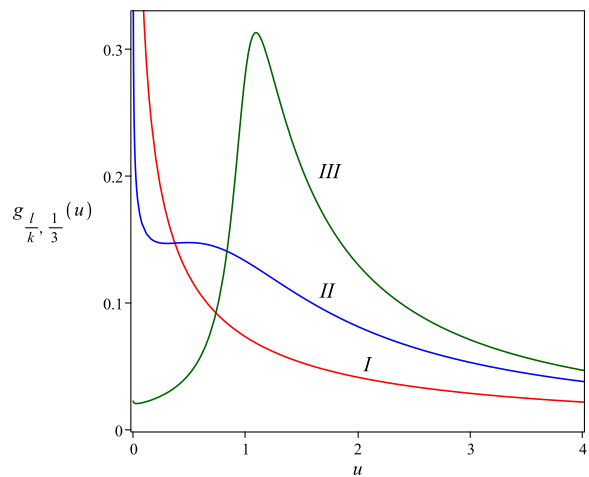}
\caption{\label{fig1}(Color online) Plot of $g_{\alpha, \beta}(u)$ given by Eq. \eqref{5/06-2} for $\beta = 1/3$ and $\alpha = 1/2$ (I; red), $\alpha = 3/4$ (II; blue), and $\alpha = 19/20$ (III; green).}
\end{figure}

Properties of $g_{l/k, \beta}(u)$, for $\beta$ fixed, are illustrated on Fig. \ref{fig1} where we show the probability densities $g_{l/k, \beta}(u)$ given by Eq. \eqref{5/06-2} for $\beta = 1/3$ and $\alpha~=~1/2, 3/4, 19/20$.
Similarly to the properties of CC relaxation Fig. \ref{fig1} shows that for fixed value of $\beta$ there exists $\alpha_{0}$ for which $g'_{\alpha_{0}, \beta}(u)|_{u=u_{0}} = 0$. For example, if $\beta = 1/3$ we get $\alpha_{0} \approx 0.747$ and $u_{0}  \approx 0.355$. If $\alpha > \alpha_{0}$ the maximum becomes higher and moves towards larger values of $u$. 

Our last step in this Section is to find the series representation of $g_{\alpha, \beta}(u)$ for $\alpha = l/k$. To realize this goal we use the series representation of the generalized hypergeometric functions, see Eq. (7.2.3.1) on p. 437 of \cite{APPrudnikov-v3}. In such a way we get in Eq.\eqref {5/06-2} a double sum: one over $r$ ($r=0, 1, 2, \ldots$) which comes from the series representation of ${_{p}F_{q}}$, and the second one over $j$ ($j = 0, 1, \ldots, k-1$) which appears in the Eq. \eqref{5/06-2} itself. Changing the summation index $k r + j \to r$ we arrive at the expression 
\begin{equation}\label{6/06-4}
g_{\alpha, \beta}(u) = \frac{1}{\pi} \sum_{r=0}^{\infty} \frac{(-1)^{r}}{r!} \frac{ (\beta)_{r} \sin[\alpha(\beta+r) \pi]}{u^{1+\alpha(\beta+r)}},
\end{equation}
which, after representing the sine function as the imaginary part of $e^{i\pi \alpha (\beta+r)}$, using the integral representation of the $\Gamma$ function \cite[Eq. (8.312.2)]{Gradshteyn07}, and applying Eq. \eqref{5/09-3}, leads us to the integral form of Eq. \eqref{6/06-4}:
\begin{equation}\label{6/06-5}
g_{\alpha, \beta}(u) = \frac{1}{\pi} {\rm Im}\left\{\int_{0}^{\infty} e^{-u\xi} F_{\alpha, \beta}(\xi e^{i\pi}) d\xi\right\}. 
\end{equation}
where the "auxiliary" function $F_{\alpha, \beta}(x)$, written down in terms of the (three-parameter) generalized Mittag-Leffler function, is 
\begin{equation}\label{11/06-2}
F_{\alpha, \beta}(x) = x^{\alpha\beta} E^{\beta}_{\alpha, 1+\alpha\beta}(-x^{\alpha}).
\end{equation}
From Eqs. \eqref{5/09-3} and \eqref{11/06-2} it is easy to demonstrate the properties of $F_{\alpha, \beta}(x)$: (i) $F_{\alpha, \beta}(x)$ vanishes at $x=0$; (ii) $F_{\alpha, \beta}(x)$ goes to $1$ in the limit of $x\to \infty$. This asymptotic behavior was found in a recent paper \cite{FMainardi15} and is given therein by an unnumbered formula on the p. 76 using slightly different notation $\alpha_{\rm M} = \alpha$, $\beta_{\rm M} = \alpha\beta + 1$ and $\gamma_{\rm M} = \beta$. 

For the CC relaxation, i.e. $\beta=1$, $F_{\alpha, \beta}(x)$ can be expressed via the classical Mittag-Leffler function, namely $F_{\alpha, 1}(x) = 1 - E_{\alpha}(-x^{\alpha})$, as quoted in \cite[Eq. (9) on p. 185]{KWeron96}. For the CD relaxation ($\alpha =1$) we have $F_{1, \beta}(x) = x^{\beta} E_{1, 1+\beta}^{\beta}(-x) = 1 - \Gamma(\beta, x)/\Gamma(\beta)$ \cite{RCFAE15}, where $\Gamma(\beta, x)$ is the incomplete gamma function. Moreover, $\ulamek{d}{d x}F_{\alpha, \beta}(x) = f_{\alpha, \beta}(x)$ which results from Eq. (1.9.6) on p. 46 of \cite{AAKilbas}.

Applying \cite[Eq. (2.5)]{TRPrabhakar71} to Eq. \eqref{6/06-5} with Eq. \eqref{11/06-2} put in and employing de Moivre's formula to calculate the imaginary part we rederive the function $g_{\alpha, \beta}(u)$ obtained and extensively studied in the just quoted \cite{FMainardi15}. Namely, for $0 < \alpha < 1$ and $\beta >0$, we get two solutions
\begin{equation}\label{14/06-2}
g_{\alpha, \beta}(u) = \pm\frac{1}{\pi u} \frac{\sin\left[\beta\theta_{\alpha}(u) \right]}{[u^{2\alpha} + 2u^{\alpha}\cos(\pi\alpha) + 1]^{\beta/2}}
\end{equation}
where 
\begin{equation}\label{15/07-1}
\theta_{\alpha}(u) = \arctan\left(\frac{u^{\alpha}\sin(\pi\alpha)}{u^{\alpha} \cos(\pi\alpha) + 1}\right),
\end{equation} 
and the sign in Eq. \eqref{14/06-2} depends on the choice of the branch of arctan function having in  Eq.\eqref{15/07-1} the essential singularity for $u =[\cos(\pi\alpha+\pi)]^{-1/\alpha}$. To preserve the physical interpretation of the function $g_{\alpha, \beta}(u)$ it has to be nonnegative. The denominator in Eq. \eqref{14/06-2} is always positive so $g_{\alpha, \beta}(u)$ remains nonnegative if $\beta\theta_{\alpha}\in[0,\pi]$ mod $2n\pi$. Because Eq.\eqref{15/07-1} gives $\theta_{\alpha}\in [0, \pi\alpha]$ this leads to $\beta<1/\alpha$.
This fully agrees with the observation made in \cite[Eq. (8)]{FMainardi15}.

\section{Relaxation function and evolution equation}

The relaxation function $n(t)/n_{0}~\equiv~[n(t)/n_{0}]_{\alpha, \beta}$ may be obtained in its explicit and exact form by calculating the Laplace transform \eqref{4/06-1} with $g_{\alpha, \beta}(u)$ given by Eq. \eqref{5/06-1}. In what follows we use the formulas from \cite{APPrudnikov-v3}, namely Eq. (2.24.3.1) on p. 350; Eq. (8.2.2.14) on p. 619, and Eq. (8.2.2.3) on p. 618 and next compare the results with Eqs. \eqref{11/06-2} and \eqref{4/06-9}. The result is \cite[Eq. (30)]{RCFAE15}:
\begin{align}\label{10/06-1}
\left[\frac{n(t)}{n_{0}}\right]_{\alpha, \beta} = 1 - F_{\alpha, \beta}(\ulamek{t}{\tau_{0}}),
\end{align}
where $F_{\alpha, \beta}(t/\tau_{0})$ is given by Eq. \eqref{11/06-2} with Eq. \eqref{4/06-9} inserted. 
\begin{figure}[!h]
\includegraphics[scale=0.40]{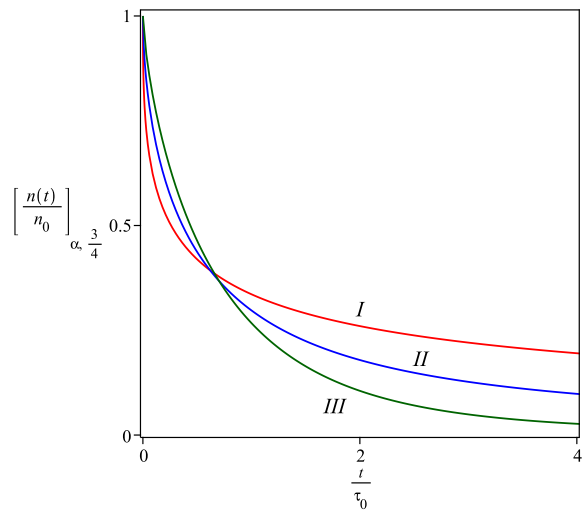}
\caption{\label{fig3}(Color online) Plot of $[n(t)/n_{0}]_{\alpha, \beta}(x)$ given in Eqs. \eqref{10/06-1} and \eqref{4/06-9} for $\beta = 3/4$ and $\alpha = 1/2$ (I; red), $\alpha = 3/4$ (II; blue), and $\alpha = 19/20$ (III; green).}
\end{figure}
As it should be, and as seen from the  Fig. \ref{fig3},  
 the rhs of Eq. \eqref{10/06-1}  is equal to 1  for $t=0$ and vanishes for $t$ going to infinity. Eq. \eqref{10/06-1} resembles the series representation of the relaxation function obtained in \cite{RHilfer02, RHilfer02a}. This confirms the hypothesis that the HN relaxation could be explained using the Debye relaxation, analogously to the particular case $\beta=1$ i.e. CC relaxation \cite{KWeron96}.

The explicit form of relaxation function $[n(t)/n_{0}]_{\alpha, \beta}$ given by Eqs. \eqref{10/06-1} and \eqref{4/06-1} enable us to derive the self-similar properties of $g_{\alpha, \beta}(u)$. Rewriting the first equality in Eq. \eqref{4/06-1}, where instead of $t/\tau_{0}$ we take $a^{1/\alpha} p$, $a > 0$, and taking into account Eq. \eqref{10/06-1}, we get
\begin{align*}
1- &(a p^{\alpha})^{\beta} E^{\beta}_{\alpha, 1 + \alpha\beta}(- ap^{\alpha}) =\! \int_{0}^{\infty}\! e^{- p a^{\frac{1}{\alpha}} u} g_{\alpha, \beta}(u) du \\
&\qquad\stackrel{u = a^{-1/\alpha} x}{=}\! \int_{0}^{\infty}\! e^{-p x} a^{-1/\alpha} g_{\alpha, \beta}(a^{-1/\alpha}x) dx, 
\end{align*}
which means that the single variable function $g_{\alpha, \beta}(u)$ can be uniquely extended to a two-variable 
one: 
\begin{equation}\label{16/09-2}
\tilde{g}_{\alpha, \beta}(a, x) = a^{-1/\alpha} g_{\alpha, \beta}(x a^{-1/\alpha}).
\end{equation}
We emphasize that Eq. \eqref{16/09-2} is a self-similarity property obeyed also by the classical {(one-parameter)} Mittag-Leffler function \cite{AAKilbas} and the L\'{e}vy stable distribution \cite{KAPenson10}. From Eq.\eqref{16/09-2} and the second equality in Eq. \eqref{4/06-1} with $a = (t/\tau_{0})^{\alpha}$, $p=1$, and for $t_{0} \leq t_{1} \leq t_{2}$ we get the Laplace-like convolution property
\begin{multline}\label{17/07-1}
\int_{0}^{x} g_{\alpha, \beta}((\ulamek{t_{2}-t_{1}}{\tau_{0}})^{\alpha}, x-y) g_{\alpha, \beta}((\ulamek{t_{1}-t_{0}}{\tau_{0}})^{\alpha}, y) dy \\= g_{\alpha, \beta}((\ulamek{t_{2}-t_{0}}{\tau_{0}})^{\alpha}, x),
\end{multline}
being a kind of the evolution equation written in the integral form. The similar rule is fulfilled by the one-sided L\'{e}vy stable distribution, see \cite[Eq. (12)]{GDattoli14} and \cite[Eq. (13)]{KGorska16}. We note that Eq. \eqref{17/07-1} differs from the standard Laplace convolution of functions depending on  \textit{one} variable. Here we do have the integral form of evolution equation for the distribution $g_{\alpha, \beta}((\ulamek{t}{\tau_{0}})^{\alpha}, x)$ depending of  \textit{two} variables where \textit{both} variables change simultaneously.



\section{$f_{\alpha, \beta}(t/\tau_{0})$ and $\varPhi_{\alpha}(t)$}

\begin{figure}[!h]
\includegraphics[scale=0.40]{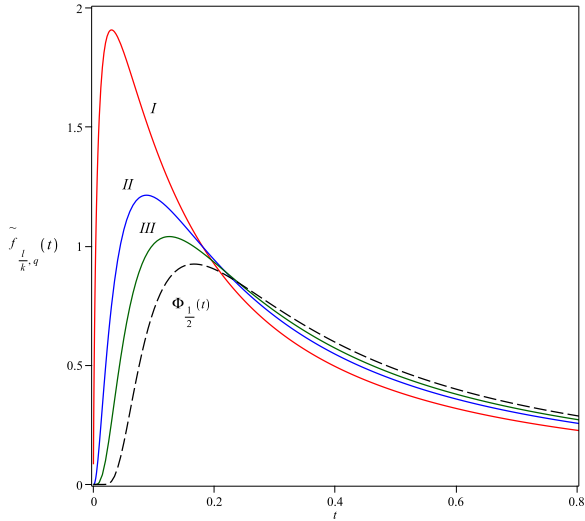}
\caption{\label{fig4}(Color online) Plot of $\tilde{f}_{1/2, q}(t)$, where the Prabhakar function Eq. \eqref{26/09-1} is defined for $\tau_{0} = (q-1)^{1/\alpha}$, $\beta=(2-q)/(q-1)$, and $q = 1.2$ (I; red), $q = 1.1$ (II; blue), and $q = 1.05$ (III; green). The black, dashed line shows the one-sided L\'{e}vy stable distribution for $\alpha = 1/2$, i.e. $\varPhi_{1/2}(t) = \exp[-1/(4t)]/(2\sqrt{\pi} t^{3/2})$.} 
\end{figure}

Following \cite{KWeron96} the relation between the HN relaxation and one-sided L\'{e}vy density $\varPhi_{\alpha}(u)$ is usually understood as \cite[Eq. (7)]{KWeron96} or  \cite[Eq. (11)]{KGorska12}. Here we propose a new kind of the connection linking $f_{\alpha, \beta}(t/\tau_{0})$ with $\varPhi_{\alpha}(t)$.
To push forward our approach we reparameterize the Eq. \eqref{3/06-2}, namely we put  $\beta = (2-q)/(q-1)$ (condition $1 < q < 2$ assures that $\beta > 0$), $\tau_{0} = (q-1)^{1/\alpha}$ and $i\omega = \kappa$. 
The  derivative (with minus sign) of Eq. \eqref{3/06-2} with respect to  $\kappa$ is the probability density function 
\begin{equation}\label{19/09-1}
W_{\alpha, q}(\kappa) = \alpha (2-q)\kappa^{\alpha-1}[1-(1-q)\kappa^{\alpha}]^{\frac{1}{1-q}}.
\end{equation}
It is called the $q$-Weibull distribution \cite{SPicoli09} and for $q\to 1+$ tends to the Weibull distribution \cite{SPicoli09, JLDevore} being the derivative of the stretched exponential. Using Eqs. \eqref{5/09-2} 
and Eq. \eqref{4/06-9} we compare $\varPhi_{1/2}(t)$ with
$\tilde{f}_{\alpha, q}(t) \equiv f_{\alpha, (2-q)/(q-1)}(t (q-1)^{-1/\alpha})$ for $\alpha = 1/2$ and $q=1.2, 1.1, 1.05$. The plot, see Fig. \ref{fig4}, strongly suggests that $\tilde{f}_{1/2, q}(t)$ goes over to the one-sided L\'{e}vy stable distribution $\varPhi_{1/2}(t)$ when $q$  approaches 1. 

The above is not accidental  -  the limit $q\to 1+$ of $\tilde{f}_{\alpha, q}(t)$ may be rigorously analyzed using Eq.\eqref{5/09-2} and the series representation of the (three-parameter) generalized Mittag-Leffler function given in Eq. \eqref{5/09-3}. After applying the Gauss-Legendre multiplication formula to the Gamma functions in the denominator of Eq. \eqref{5/09-3} we get
\begin{align}\label{19/09-2}
\begin{split}
\tilde{f}_{\alpha, q}(t) &= \sum_{n=0}^{\infty} \frac{(-1)^{n}}{n!} \big(\ulamek{2-q}{q-1}\big)_{n} \left[\ulamek{t}{(q-1)^{-1/\alpha}}\right]^{\alpha(n + \frac{2-q}{q-1}) - 1} \\
& \times \Gamma[1-\alpha(n+\ulamek{2-q}{q-1})]\, \sin[\alpha(n+\ulamek{2-q}{q-1})].
\end{split}
\end{align}
Writing down the sine function as the imaginary part of $\exp[i\pi\alpha(n+\ulamek{2-q}{q-1})]$ and employing Eq. (8.312.2) on p. 892 of \cite{Gradshteyn07}, one gets the integral representation of $\tilde{f}_{\alpha, q}(t)$ 
\begin{equation}\label{19/09-3}
\tilde{f}_{\alpha, q}(t) = {\rm Im} \left\{\int_{0}^{\infty} e^{-ty}[1+(q-1)(ye^{-i\pi})^{\alpha}]^{-\frac{2-q}{q-1}} \frac{dy}{\pi} \right\},
\end{equation}
which for $q\to 1+$ goes to the one-sided L\'{e}vy stable distribution $\varPhi_{\alpha}(t)$ represented according to the first unnumbered equation on the last page of \cite{HPollard46}. It may be also shown that the Stieltjes moments of $f_{\alpha, \beta}(t/\tau_{0})$ given by Eq. \eqref{25/09-1} go, for $q\to 1+$, to the Stieltjes moments of $\varPhi_{\alpha}(t)$. To get this we put $\tau_{0} = (q-1)^{1/\alpha}$ and $\beta = (2-q)/(q-1)$ with $1 < q < 2$ and apply the Stirling formula for $\Gamma(\beta + \nu/\alpha)$ and $\Gamma(\beta)$ with $\beta$ as just mentioned. Moreover, the HN relaxation Eq. \eqref{3/06-2} reparametrized as above tends to the stretched exponential in the limit of $q\to 1+$.

\section{Conclusions}

Starting form the basic concepts used in the relaxation theory we have calculated the response, probability density and  relaxation functions of the HN relaxation. The response functions $f_{\alpha, \beta}(t/\tau_{0})$, given via the Prabhakar functions and related to the (three-parameter) generalized Mittag-Leffler functions, have been found to be given for rational $\alpha$ as finite sums of the generalized hypergeometric functions. We have also expressed the (three-parameter) generalized Mittag-Leffler functions via the one-sided L\'{e}vy stable distributions $\varPhi_{\alpha}(t)$. In such a way we generalized the relation between the classical (one-parameter) Mittag-Leffler function and the one-sided L\'{e}vy stable distribution known from the case $\beta = 1$, i.e.  the CC relaxation. Taking into account the evidence  that $f_{\alpha, \beta}(t/\tau_{0})$ and the appropriate $\varPhi_{\alpha}(t)$ obey the same asymptotic behavior, we have suggested the existence of a deeper  (physical) relation between $f_{\alpha, \beta}(t/\tau_{0})$ and $\varPhi_{\alpha}(t)$, to be investigated in the future. We have also identified the values of $\beta$, namely $0 < \beta \leq 1/\alpha$, for which the normalizable function $g_{\alpha, \beta}(u)$, connected to the response function via its inverse Laplace transform, may be considered as a probability density which higher moments 
do not exist. 
For probability densities $g_{\alpha, \beta}(u)$  we have also derived the Laplace-like convolution relations which may be interpreted as integral forms of the evolution equations. 

In our opinion the main benefit of the paper is, besides providing the explicit and exact construction of the HN relaxation-related functions, that we were able to represent them as well-known special functions which are routinely implemented in the standard computer algebra systems. The use of generalized hypergeometric functions simplifies calculations and allows to obtain many valuable results in a relatively simple and quick way.  Consequently, we claim that computational methods developed in our paper help to shed light how various versions of $f_{\alpha, \beta}(t/\tau_{0})$, $g_{\alpha, \beta}(u)$, and $[n(t)/n_{0}]_{\alpha, \beta}$ appearing in the literature can be connected each to another and what is their relation to the L\'{e}vy distributions. The latter is evidently seen mathematically and seems to be relatively easy to understand as a mathematical fact but the physical background and meaning of such a relation  needs careful explanation. In particular, one needs to investigate a possible existence of common sources of anomalous diffusion and non-Debye relaxation patterns and/or the emergence of dynamics governed by fractional equations.
  
\section*{Acknowledgments}
K. G., A. H. and K. A. P. were supported by the PAN-CNRS program for French-Polish collaboration. Moreover, K. G. thanks for support from MNiSW (Warsaw, Poland), "Iuventus Plus 2015-2016", program no IP2014 013073.

\end{document}